# Fully 3D-Printed Wideband Metasurface Folded Reflectarray Antenna

Evangelos Vassos, Thomas Whittaker, Abdul Jabbar, Aakash Bansal, and Will Whittow

*Abstract*—This article presents a fully 3D-printed wideband metasurface folded reflectarray antenna (MFRA) operating in the millimeter-wave n257 band. The proposed MFRA integrates a novel polarization-rotating reflective metasurface (RMS), a compact embedded horn feed, and a polarization-selective metasurface polarization grid (MPG), all fabricated using a low-cost in-house 3D-printed method. Unlike conventional PCB-based FRAs constrained to planar unit-cell geometries, the proposed anisotropic meta-element design exploits full three-dimensional dielectric control by tailoring varying unit-cell heights. This volumetric tuning, combined with the spatial distribution of the meta-elements, enables phase compensation exceeding 400° across the aperture, supporting robust wideband performance. An MFRA prototype is in-house fabricated and experimentally validated. Measured results agree well with simulations, achieving a −10 dB impedance bandwidth of 20.7% (26–32 GHz) and a peak realized gain of 31.1 dBi at 28.2 GHz. The antenna exhibits sidelobe levels below −20 dB, cross-polarization below −30 dB, and a compact height-to-diameter ratio of 0.20. Stable pencil beams with an average HPBW of 3.7° are maintained across the operating band. To further validate the robustness of the proposed in-house designed MFRA, a commercially manufactured RMS was also obtained, whose measured performance shows excellent agreement with the in-house 3D-printed version, confirming a cost-effective rapid-prototyping antenna solution. The proposed MFRA is a cost-effective solution for beyond 5G and 6G high-gain point-to-point mmWave wireless applications, such as fixed wireless access, near field communication, and beam focusing.

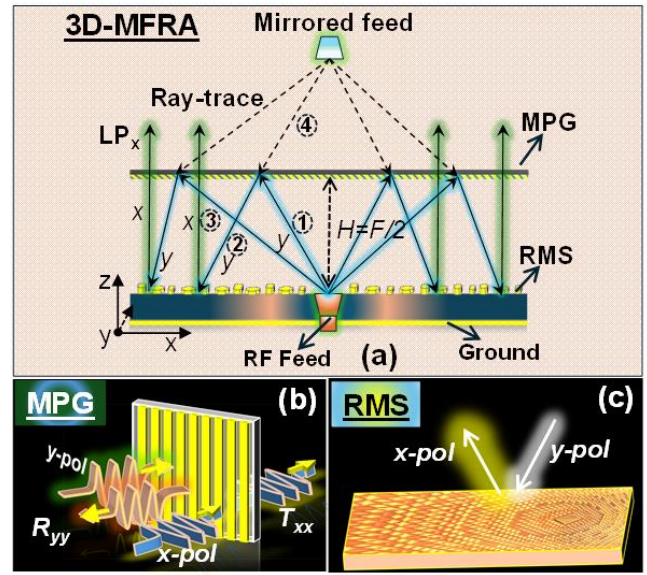

Fig. 1. (a) Schematic design and working principle of the proposed MFRA. (b) Illustration of the working principle of the designed MPG and (c) RMS.

## I. Introduction

METAMATERIALS and their two-dimensional (2D) planar counterparts, metasurfaces (MSs), have revolutionized the landscape of application of electromagnetic (EM) waves since their conception and inception [1], [2]. This is due to the unprecedented ability of specifically designed MSs to control the amplitude, phase, or polarization of the EM wave through the collective response of their subwavelength constituent unit-cell geometries, which otherwise is not possible in naturally occurring materials.

As demand for high-throughput 6G and beyond wireless networks increases, high-directive, low-profile antenna arrays become crucial as the operating frequency increases. To meet high-gain demands and reduce the size of traditional reflectarray antennas, folded reflectarray antennas (FRAs) were introduced, embedding the feed within the antenna structure and folding the propagation path between a main reflective surface and a polarization-selective layer. [3], [4].

Contemporary FRA implementations are predominantly realized using printed circuit board (PCB) technology due to ease of design conception, and fabrication maturity [5]–[11]. However, as aperture size increases, PCB-based FRAs face growing dielectric and conductor losses, fabrication cost, and scalability constraints. Apart from traditional PCB fabrication technology, the concept of growing dielectric structures in three-dimensional (3D) space offers a more flexible and versatile metasurface antenna design approach [12]. This synthesizing method is widely known as additive manufacturing or 3D printing [13], [14]. In particular, fused deposition modelling (FDM) is one of the low-cost additive

This work was supported in part by SYnthesizing 3D METAmaterials for RF, microwave and THz applications (SYMETA) under EPSRC Grant EP/N010493/1, in part by Anisotropic Microwave/Terahertz Metamaterials for Satellite Applications (ANISAT) under Grant EP/S030301/1, and in part by Transparent Transmitters and Programmable Metasurfaces for Transport and Beyond-5G (TRANSMETA) under Grant EP/W037734/1. (*Corresponding author: Abdul Jabbar*.)

The authors are with Wolfson School of Mechanical, Electrical and Manufacturing Engineering, Loughborough University, UK (e-mail: evangelos.vassos@steatite.co.uk;{t.whittaker;a.jabbar;a.bansal;w.g.whittow}@lboro.ac.uk).

manufacturing techniques with dielectric materials to attain complex 3D geometries, enabling fast and low-cost prototyping [15], [16]. It has gained attention for the rapid prototyping of microwave and mmWave antennas, MSs, and components [17]–[22]. Although additively manufactured reflectarray antennas have attracted increasing attention [23]–[27], most reported designs remain limited to spatially fed or PCB-based configurations. Existing research in reflectarray additive manufacturing focuses on planar or perforated dielectric structures, but not on fully 3D-printed FRAs with integrated feed and volumetric metasurface control. Some 3D-printed FRA architectures [28]–[30] have been presented in the literature; however, achieving low-profile, high-gain, and wide-bandwidth at mmWave bands with fully 3D printed designs remains limited in the literature. This underscores the exploration of the development of fully additively manufactured mmWave FRAs capable of delivering robust electromagnetic performance.

*A. Novelty and Contribution*

In this work, we propose a novel fully additively manufactured folded metasurface folded reflectarray antenna (MFRA) to operate around 28 GHz mmWave band. The antenna consists of three main components: a newly designed wideband anisotropic reflective metasurface (RMS) at the bottom, a top metasurface polarizing grid (MPG), and an integrated horn antenna as a feeding source (see Fig. 1(a)). All three antenna components are fully 3D-printed.

The designed RMS can reflect a spherical wave into a cross-polarized plane wave, while the MPG offers polarization-selective behaviour by reflecting one LP wave while transmitting the other orthogonal one. Consequently, a highly directive collimated beam can be generated in the far-field region. A fully 3D-printed MFRA antenna integrating these two types of MSs and an LP feeding source is designed, fabricated, and measured. Both simulated and measured results show that the proposed MFRA can achieve a wide bandwidth covering 25 to 32 GHz, a compact height-to-diameter ($H/D$) ratio of 0.20, and a measured realized gain of 31.1 dBi at 28 GHz.

Although the proposed proof-of-concept MFRA prototype is demonstrated for linear polarization, the modular and detachable RMS architecture provides flexibility for future integration of circularly polarized metasurfaces around the 28 GHz band, to offer extension to CP operation if required. In addition, the use of a standard, integrable feeding interface allows the MFRA to be excited by a wide range of third-party mmWave horn antennas, validating its off-the-shelf compatibility and practical deployment feasibility.

Furthermore, two RMS prototypes were fabricated, one produced in-house using low-cost FDM 3D printing with a HIPS dielectric, and the other commercially manufactured using Nano Dimension's advanced DragonFly IV 3D printing system [31]. Both prototypes exhibit closely matched measured gain (within 1 dB variation) and impedance bandwidth. This close agreement demonstrates the robustness of the proposed design and confirms that it can be rapidly realized using inexpensive materials and accessible 3D-printing techniques, hence reducing fabrication cost by hundreds of dollars while significantly shortening the lead time.

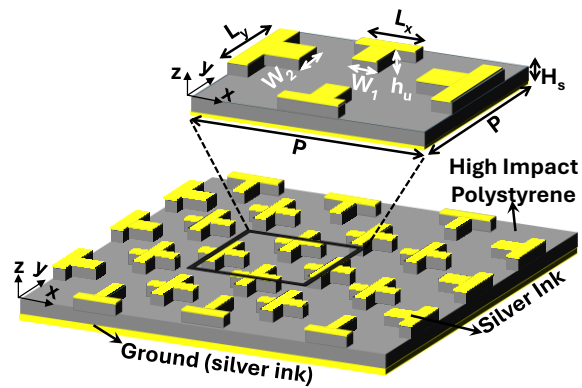

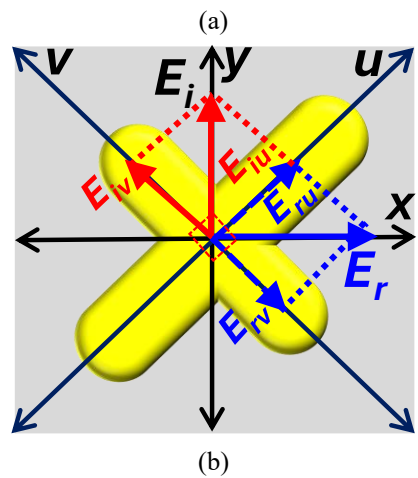

Fig. 2. (a) Geometry of the unit cell lattice of RMS. The dimensions are as follows: $P = 6$, $H_s = 0.4$, $W_1 = 6$, $W_2 = 6$, $h_u = 0.2$ and 0.4, attained values for $L_x$ and $L_y$ = 0.1, 0.4, 1.7, 2.6, 2.9, 3.3, 3.4, 3.5, 3.6, 4.1, 4.8, 5.2 (units: mm). (b) Polarization conversion of a y-polarized field into an x-polarized field, the y-polarized electric field is resolved into two orthogonal eigen vector components along the u- and v-axis.

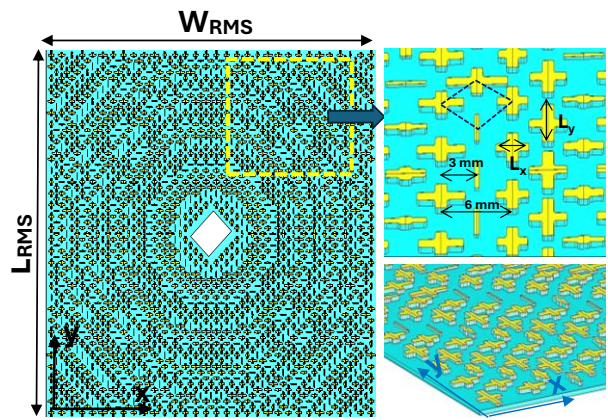

Fig. 3. Schematic of the complete triangular lattice RMS simulated in CST, with a zoomed-in section and a perspective cross-sectional view. $L_{RMS} = W_{RMS} = 192$ mm.

## II. OPERATING PRINCIPLE OF MFRA

The architecture and the radiation principle of the proposed MFRA are illustrated in Fig. 1(a). As depicted, it comprises three main components, including an RMS, an MPG, and an LP feeding source. As depicted in Fig. 1(b), the top MPG functions as a polarization-selective surface, reflecting the *y*-polarized wave with high efficiency, while transmitting the orthogonal *x*-polarized wave. The RMS provides a 90° rotation of polarization, as shown in Fig. 1(c). Based on the above characteristics, the operation of the proposed MFRA can be described as follows:

A *y*-polarized spherical wave (ray1) radiated by the LP feed first impinges on the top MPG and is reflected back without a change in polarization (ray2) toward the RMS. Then, the reflective MS performs 90° polarization rotation and converts the y-polarized incident wave into an x-polarized wave while collimating a high-gain beam with low cross-polarization. Finally, the collimated x-polarized wave passes through the MPG without any change in polarization.

According to mirror-image theory, the folded ray path between the MPG and RMS is equivalent to a conventional reflectarray illuminated by an imaginary feed at the focal ($F$) point [11]. As the electromagnetic wave undergoes two controlled specular reflections within the folded structure, the effective propagation path is preserved while the F/D profile of the antenna is significantly reduced. The incident wave emitted from the feeding source is reflected twice between MPG and RMS. According to the folded ray tracing path illustrated in Fig. 1(a), we can achieve:

$$ray1 + ray2 = ray4 + ray2 \quad (1)$$

where $ray1 = ray2 = ray4$. On this basis, the profile height of MFRA can be decreased to about half the focal length. Therefore, the proposed MFRA offers a compact profile (with reduced F/D and hence a lower *H/D* ratio).

## III. DESIGN AND ANALYSIS OF MFRA SYSTEM

### A. Design and Analysis of Unit Cell for RMS

The bottom MS, referred here as RMS, is the key component of the proposed MFRA. The schematic diagram of the unit cell lattice is shown in Fig. 2(a). A cross-shaped parallelepiped geometry was selected for the unit cell design. The chosen dielectric material is High Impact Polystyrene (HIPS), having a thickness of 0.4 mm, a dielectric constant of 2.49, and a loss tangent of 0.00084. The bottom side of the dielectric is fully grounded using conductive silver ink of conductivity $1\times10^5$ S/m, and thickness 0.01 mm, to reflect back EM signal.

In addition to the width ($L_x$) and length ($L_y$) of the unit cells, a third degree of freedom, the volumetric height ($h_u$), is also controlled in the proposed design. Two different values for $h_u$, 0.2 and 0.4 mm, are utilized across the MS. This variation in unit-cell elevation provides an additional volumetric degree of freedom enabled by additive manufacturing, as depicted in Fig. 3, leading to additional resonances in the desired band of interest to offer a wideband response. It is noteworthy that such a degree of freedom in volumetric control of unit cells is not possible in planar PCB-based metasurfaces, where only the length and width of a unit-cell patch can be controlled.

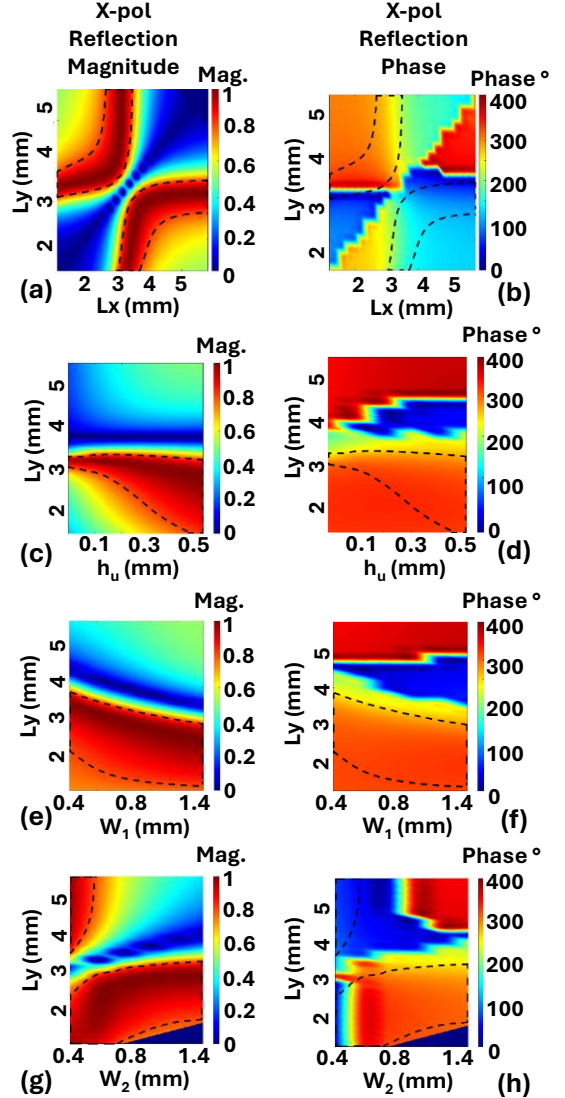

Fig. 4. Magnitude and phase profile of cross-polarized (X-pol) reflection coefficient of the proposed unit cell lattice for various geometric combinations. Heat maps (a,c,e,g) show reflection magnitude for varying $L_y$ across $L_x$, $h_u$, $W_1$, and $W_2$, respectively. Similarly, graphs (b, d, f, h) reveal the phase profile.

Since the RMS is backed by a continuous ground plane, transmission is suppressed, and the design focuses on maximizing the off-diagonal reflection terms, $R_{xy}$ and $R_{yx}$, to realize y→x or x→y polarization conversion [9]. For y→x conversion for *y*-incident $E_y^{in}$, or vice versa for $E_x^{in}$, $|R_{xy}| \approx |R_{yx}| \approx 1, and\ |R_{yy}| \approx 0$.

To further elaborate the cross-polarization conversion, we can determine the eigen-polarizations and eigenvalues of the proposed unit cell design. As shown in Fig. 2(b), consider a normally incident *y*-polarized electromagnetic wave $\boldsymbol{E_i} = \hat{y}E_ie^{ikz}$ having wave-number *k* striking the RMS. As shown in Fig. 2(b), there exist two coordinate systems, *xy* and *uv*, where *the u and v-axes are oriented at +45° to the x- and y-axes,* respectively. It can be observed from Fig. 2(b) that the proposed unit cell exhibits anisotropy along the *u-* and *v*-axes

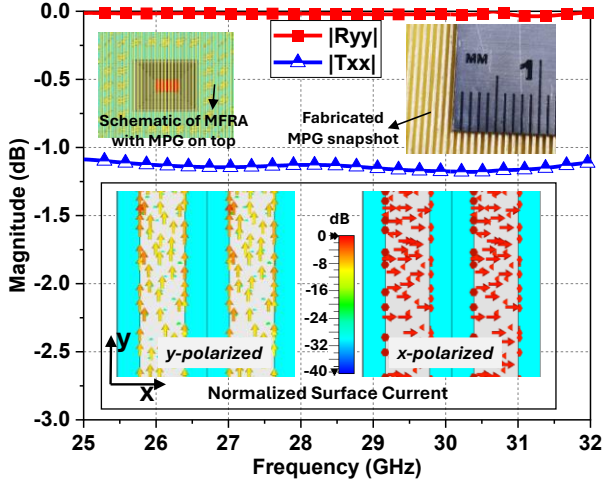

Fig. 5. Polarization-selective response of the designed MPG surface in terms of the reflection coefficient ($R_{yy}$) and transmission coefficient ($T_{xx}$). The inset figure on the top right side shows the fabricated MPG sample.

and has mirror symmetry along the $v$-axis.

The incident electric field at $z = 0$ can be resolved into two orthogonal $u$ and $v$ eigen components as (2):

$$\boldsymbol{E_i} = \hat{y}E_i = \hat{u}E_{iu}e^{i\varphi} + \hat{v}E_{iv}e^{i\varphi} \qquad (2)$$

where $E_{iu} = E_{iv} = \frac{1}{\sqrt{2}} E_i$. As $u$- and $v$-polarized components, $E_{iu}$ and $E_{iv}$, are reflected with the same magnitude, $E_{ru} = E_{rv} = E_r$. Moreover, the phase of the reflected $u$-component is $E_{ru} = 0°$ with respect to that of the incident wave $u$-component, whereas for the reflected $v$-component, it is out of phase with respect to the incident $v$-component, i.e., $\varphi_v$ of $E_{rv} = 180°$, therefore, the reflected field can be expressed as (3):

$$\boldsymbol{E_r} = \hat{u}R_u E_{iu} e^{i\varphi} + \hat{v}R_v E_{iv} e^{i\varphi} \qquad (3)$$

where $\boldsymbol{R_u} = |R_u|e^{i\varphi_u} = \frac{E_{ru}}{E_{iu}}$ and $\boldsymbol{R_v} = |R_v|e^{i\varphi_v} = \frac{E_{rv}}{E_{iv}}$ are the complex reflection coefficients. Note that if in the band of interest, $|R_u| \approx |R_v| \approx 1$ and one component of the incident wave is reflected in phase (a phase difference of $\Delta\phi \approx 0°$) with the component of the incident electric field along the same axis, while the other orthogonal component is reflected out of phase ($\Delta\phi \approx 180°$), then the electric field of the reflected wave is rotated 90° with respect to the electric field of the incident wave, as depicted in Fig. 2(b), and cross polarization conversion is achieved. As $\boldsymbol{E_r}$, obtained from the vector addition of $E_{ru}$ and $E_{rv}$, is along the $x$-axis, the reflected field is along the x-axis and can be given in the form of (4) as:

$$\boldsymbol{E_r} = \hat{u}E_r - \hat{v}E_r = \hat{x}E_r \qquad (4)$$

Hence, the incident $y$-polarized wave is reflected as the $x$-polarized wave from RMS.

The unit cell must ensure sufficient phase coverage for achieving the desired phase shift between the incident and the reflected field, and also to twist the field polarization of 90°. To design the proper element distribution on the RMS, the

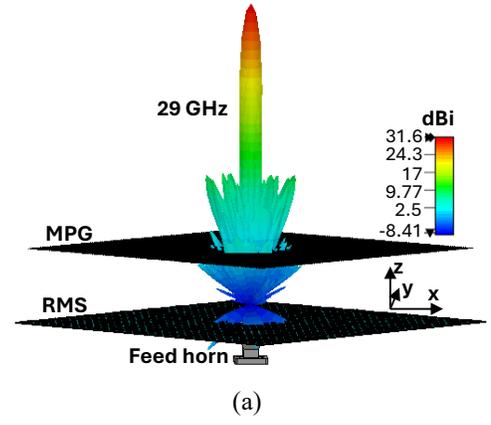

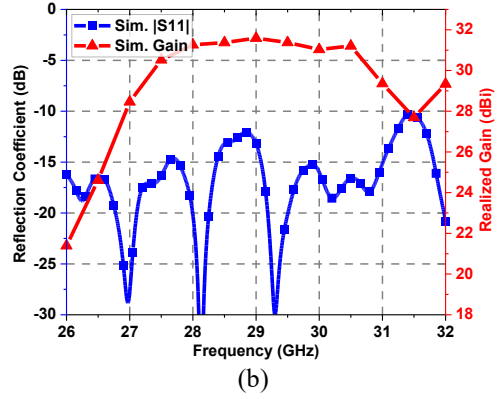

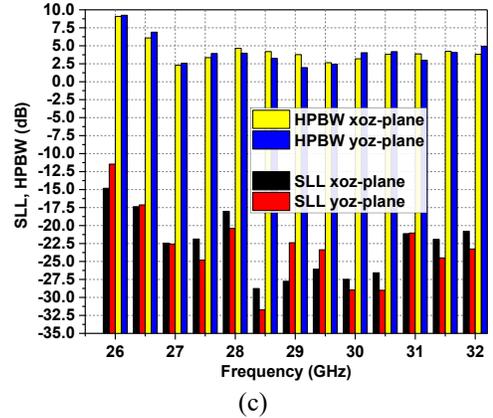

Fig. 6 (a) Schematic design of MFRA simulated in CST, showing simulated 3D gain pattern at 29 GHz. (b) Simulated reflection coefficient and realized gain of MFRA. (c) Simulated SLL and HPBW of MFRA in both xoz and yoz planes.

reflection coefficient has been calculated by varying $L_x$ and $L_y$ independently. Both $L_x$ and $L_y$ of the unit cell resonator are varied between $L_{min} = 0.1$ mm to $L_{max} = 5.2$ mm. The unit cells are arranged in a triangular lattice. The proposed 3D-printed cross-shaped unit cell exhibits anisotropic electromagnetic behaviour due to varied geometrical parameters along the orthogonal axes, enabling efficient polarization conversion ($y{\rightarrow}x$ and $x{\rightarrow}y$). Such a unit cell distribution across the RMS provides control over local EM response and the required phase profile for the reflected cross-polar field ($R_{xy}$ or $R_{yx}$).

The phase and magnitude response of the reflection coefficient of the RMS lattice are shown in Fig. 4. Parametric sweep for $L_x$ and $L_y$ demonstrates a continuous phase coverage exceeding 400°, ensuring complete $2\pi$ phase control across the aperture and a high degree of precision in phase compensation, as well as a region for maximum cross-polarized reflection magnitude, as shown in Fig. 3(a).

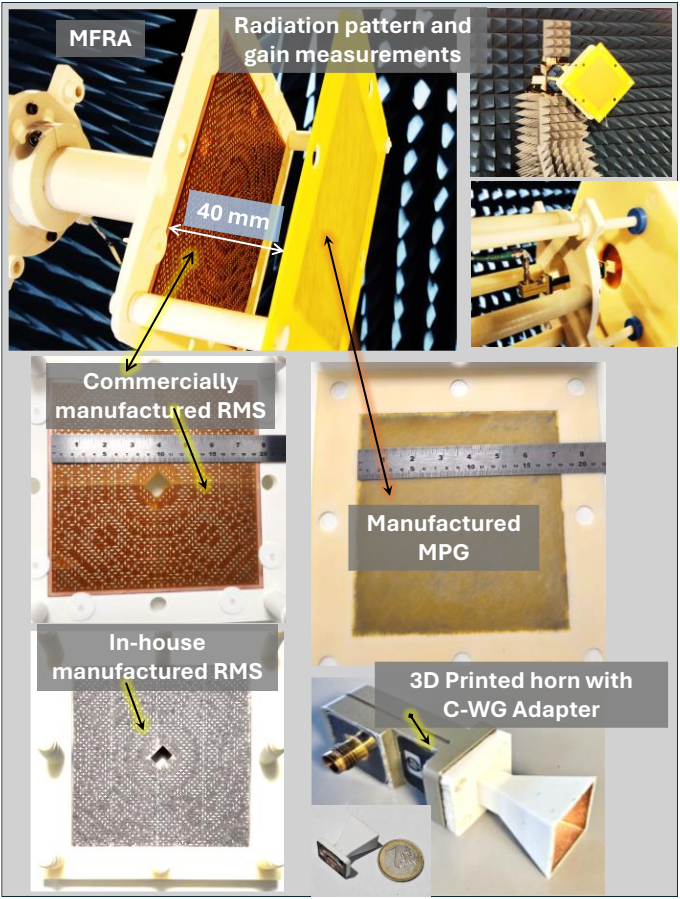

Fig. 7. Photograph of the manufactured MFRA prototype (both in-house and Nano Dimension's manufactured) alongside the measurement setup in the anechoic chamber.

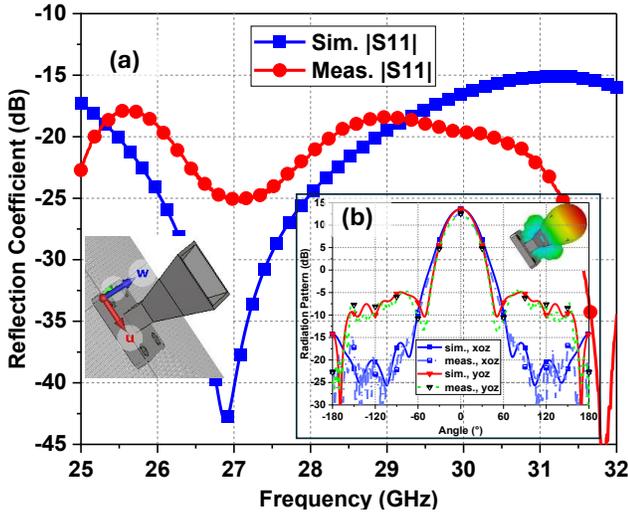

Fig. 8. (a) Measured and simulated reflection coefficient of the 3D manufactured horn antenna. (b) Simulated and measured radiation pattern at 28 GHz.

Similarly, Fig. 3(b), (c), and (d) reveal the full range of cross-polarized reflection phase and magnitude profile corresponding to $L_y$ vs. $h_u$, $L_y$ vs. $W_1$, and $L_y$ vs. $W_2$, respectively, keeping $L_x$ constant at 2.2 mm. Each of these combinations is affected so that the reflected field components present the desired phase delay and the same phase delay plus an additional 180° phase shift. Such a unit cell can be conveniently manufactured through additive manufacturing techniques, such as FDM.

Adopting the proposed unit-cell triangular-lattice symmetry introduced above, the designed full RMS structure in CST is shown in Fig. 4. The RMS structure is a square geometry, with four tiled structures arranged in four quadrants around the 45° oriented central feed aperture. The lattice periodicity ($P$) is 6 mm as indicated in Fig. 2(b), whereas the distance between any two consecutive unit cells within a lattice is 3 mm (0.28 $\lambda_0$), where in this work, $\lambda_0 = 10.17$ mm, is the free space wavelength at 28 GHz. Each tile consists of 16 × 31 unit cells; hence, in total, RMS comprises (4×(16 × 31)) = 1984 unit cells and has an area of 192 mm × 192 mm ($D$ = 192 mm = 17.92 $\lambda_0$).

### B. Design and Analysis of Top MPG Metasurface

The top MPG is designed to reflect the *y*-polarized wave, while allowing the *x*-polarized wave to pass through it. To design MPG, the same HIPS dielectric (0.4 mm thick) and a 0.01 mm silver-ink strip for metallization were used. The grid polarizer consists of periodic, parallel conductive strips arranged along one direction (assume the strips run along the *x*-axis) as shown in Fig. 5 (inset). We carried out numerical simulations in full-wave electromagnetic CST using periodic boundary conditions and Floquet ports. The width of conducting strips is 0.5 mm (0.046 $\lambda_0$). The center-to-center gap between any two consecutive strips is 1 mm (0.092 $\lambda_0$), which is much smaller than the operating wavelength. Thus, the structure behaves as an effective metasurface rather than a diffraction grating.

The complete MPG has dimensions of 180 mm × 180 mm (16.8 $\lambda_0$ × 16.8 $\lambda_0$), and comprises about 181 conducting grid unit cells. The design aims to keep *x*-polarized waves parallel to the grid lines, while *y*-polarized waves are perpendicular to them. Hence, the role of MPG is purely polarization routing.

The simulated magnitude of reflection ($R_{yy}$) and transmission response ($T_{xx}$) of MPG is shown in Fig. 5. The reflection and transmission coefficients were extracted from the fundamental Floquet mode for both TE and TM excitations, corresponding to *y*- and *x*-polarized incidences, respectively. It is observed that the magnitude of $R_{yy}$ is above -0.040 dB, and that of the $T_{xx}$ is above -0.18. This confirms that the proposed element is an excellent reflector for *y*-polarization, while allowing transmission of x-polarized waves through it with high efficiency and low loss. To illustrate the working mechanism of the proposed element, the surface current distributions are extracted and displayed in Fig. 5 (bottom inset).

For y-polarized excitation (electric field parallel to the strips), strong surface currents are induced along the metallic strips, yielding a low effective surface impedance. Strong

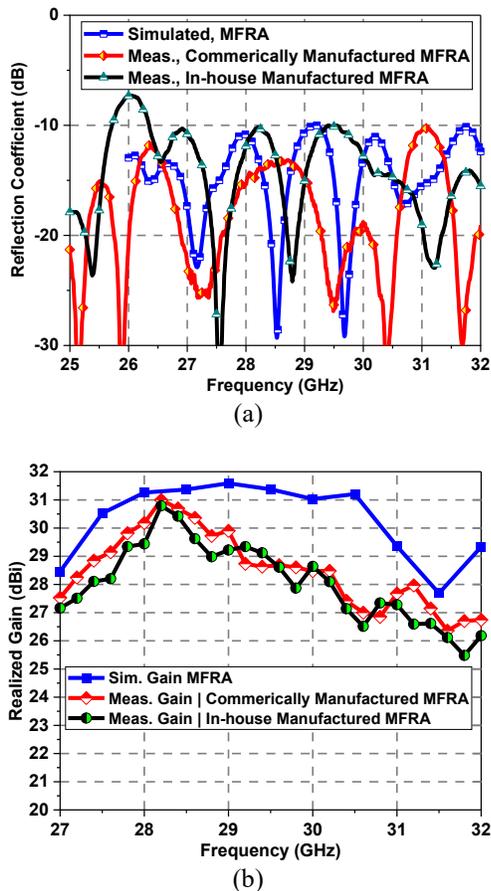

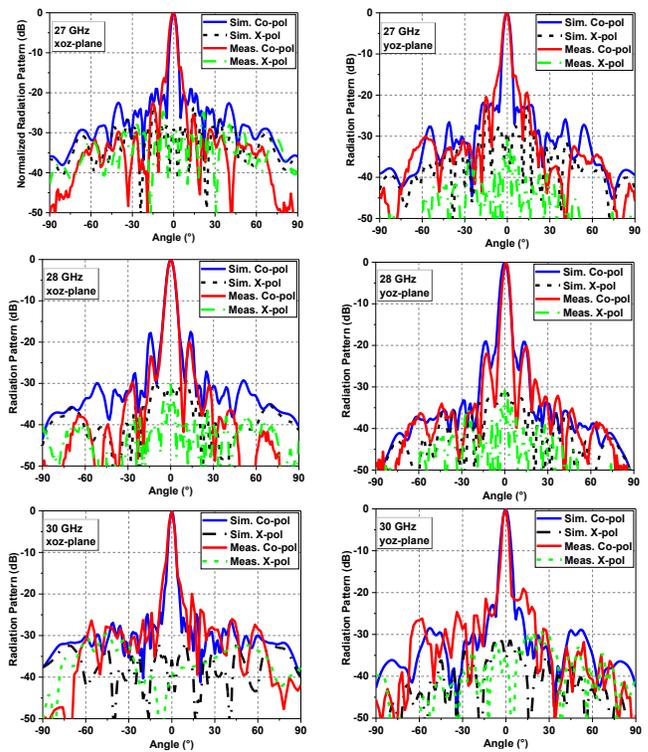

Fig. 9 (a) Measured reflection coefficient of the MFRA. (b) Measured realized gain of the MFRA.

induced surface currents cancel the tangential field, and as a result, MPG behaves like a metallic mirror, and nearly all *y*-polarized energy is reflected with high efficiency without altering its polarization.

For an *x*-polarized incident wave, the electric field is perpendicular to the grid lines. Hence, induced charges cannot flow across the grid gap because there is no continuous current path. Only weak displacement currents can be induced. As a result, the MPG exhibits a high-impedance surface effect, due to which the *y*-polarized wave interacts minimally and is transmitted through the MPG with low loss. The excellent polarization-selective performance of MPG proves its utility as a candidate for the top MS in the proposed MFRA design.

### C. Simulations of Fully Integrated MFRA

The simulated structure of the complete MFRA comprising RMS, MPG, and horn antenna, is shown in Fig. 6(a). A 45° rectangular aperture with dimensions 18.5 mm × 14.9 mm is designed at the center of RMS to integrate the horn antenna. The 45° orientation is required to efficiently excite both eigenmodes of the rotated anisotropic unit cells. The bottom of the RMS is completely painted with silver spray to serve as a ground layer for reflection.

We simulated the full MFRA structure in CST using waveguide port excitation. The optimized height (*H*) between the top MPG and bottom RMS was found to be 40 mm (3.73 $\lambda_0$), thus giving an *H/D* ratio of 0.208, and an *F/D* ratio of 0.41. The simulated reflection coefficient and realized gain of the MFRA are shown in Fig. 6(b), where -10 dB impedance is 20.69% (from 26 GHz to 32 GHz), and the simulated peak realized gain (*G*) is 31.59 dBi at 29 GHz. The 3 dB and 1 dB gain bandwidths are 14.38% and 10.67%, respectively. The radiation efficiency of the MFRA system is above 91%, while the total efficiency (including return losses) is above 78% in the 26–32 GHz band. The peak simulated aperture efficiency ($G\lambda_0^2/4\pi D$) relative to the peak gain and the square geometry (*D* = 192 mm) of the MFRA at 29 GHz is about 33.3%.

The normalized simulated radiation patterns in the xoz and yoz planes are shown along with the measured plots. However, the 3D pencil beam radiation pattern at 29 GHz can be seen in Fig. 6(a). The simulated half-power beamwidth (HPBW) remains between 2.6° and 4.5°, whereas the sidelobe levels (SLL) are below -20 dB from 27–32 GHz, as shown in Fig. 6(c). The cross-polarization (X-pol) levels remain below -30 dB. The main beam is towards broadside across the whole band of interest without exhibiting beam squint.

### IV. PROTOTYPE MANUFACTURING AND MEASUREMENT RESULTS

The manufactured prototype of MFRA is shown in Fig. 7. All MFRA parts are in-house manufactured using cost-effective FDM 3-D printing. The metallization was performed using silver conducting spray with a total metallic thickness of 0.01 mm. Four 3D-printed nylon columns serve as supporting columns between the bottom and top MSs, maintaining focal length and stability. The gap between the top and bottom MS is 40 mm, resulting in an *H/D* ratio of 0.20.

TABLE I
COMPARISON BETWEEN THE PROPOSED MFRA AND PREVIOUSLY REPORTED FRA ANTENNAS.

| Ref. | Antenna Design Technology | Center Freq. (GHz) | -10 dB BW (%) ($f_{min}$-$f_{max}$) (GHz) | FRA Size, D ($\lambda_0$) | Profile Height, H ($\lambda_0$) | H/D Ratio | Peak Gain (dBi) | 3 dB Gain BW (%) | Aperture Efficiency (%) |
|---|---|---|---|---|---|---|---|---|---|
| [5] | PCB | 26.5 | 1.51 (26.2 - 26.6) | 12.4 | 5.45 | 0.44 | 24.6 | ~2.2 | 15 |
| [6] | PCB | 30 | 5.7 (29.1 - 30.8) | 29 | 8.99 | 0.31 | 32.7 | 7.3 | 22.6 |
| [7] | PCB | 10 | 14.6 (9.5 - 11) | 10.8 | 5.4 | 0.50 | 28.2 | 15.2 | 43.3 |
| [8] | PCB | 30 | 7.2 (29.2 - 31.4) | 2.5 | 0.26 | 0.11 | 25 | 7.2 | 43 |
| [9] | PCB | 10.3 | 16.2 (9.6 - 11.3) | 7.3 | 1.17 | 0.16 | 21.9 | 11.6 | 21.8 |
| [10] | PCB | 30 | 20.3 (26.5 - 32.5) | 18 | 6.3 | 0.35 | 33.9 | 15.1 | 60.9 |
| [11] | PCB | 10 | 40 (8 - 12) | 6.4 | 1.98 | 0.31 | 24.1 | 27.5 | 51 |
| [30] | 3D printed | 17 | 17.14 (16 - 19) | 10.6 | 4.39 | 0.41 | 28.1 | 20 | 54 |
| **This Work** | **Fully 3D printed** | **28** | **20.7 (26 - 32)** | **17.9** | **3.73** | **0.20** | **31.1** | **11.2** | **31.1** |

In addition to an in-house designed prototype, we also obtained an RMS sample manufactured from Nano Dimension's DragonFly IV advanced 3D printing system, to compare the in-house-designed performance with that of the commercially manufactured one.

A 3D-printed lightweight (less than 2 grams) feed horn antenna is attached to the designated aperture at the bottom of the RMS for excitation. We simulated and designed a horn antenna to cover the desired mmWave bandwidth and fabricated it using additive manufacturing. To measure the horn antenna, a Keysight R281A (2.4 mm standard, 26 to 40 GHz, 50 Ohm) coaxial waveguide adapter was used. Simulated and measured results of the horn show a good match, and the magnitude S11 is well below -15 dB between 25–32 GHz, as shown in Fig. 8(a). The measured and simulated radiation patterns of the separate horn antenna are also characterized, showing an excellent match. The xoz and yoz plane patterns at 28 GHz are shown in Fig. 8(b). The HPBW is about 40° and 36° in the *xoz* and *yoz* planes, respectively. The simulated and measured realized gain of the horn lies between 13.28 and 13.66 dBi and between 12.05 and 13.26 dBi, respectively, within 27–30 GHz.

The reflection coefficient of both MFRA samples was measured on an Anritsu MS46522B VNA. As shown in Fig. 9(a), the measured -10 dB impedance bandwidth of both MFRA prototypes covers 20.69%, ranging from 26 to 32 GHz, and agrees well with the simulation result. The reflection coefficient is very stable and repeatable when measured through multiple trials in different antenna positions.

The gain and radiation patterns were measured in the anechoic chamber. The peak measured broadside gain is 31.1 dBi at 28.2 GHz, as shown in Fig. 9(b). Note that despite the significant difference in fabrication cost and processing methods, both prototypes exhibit closed measured gain, with a variation of less than 1 dB. This close agreement demonstrates the robustness of the proposed design and confirms that it can be rapidly realized using inexpensive materials and accessible 3D-printing techniques. Consequently, it reduces fabrication cost by hundreds of dollars while significantly shortening the prototyping lead time.

The measured 3dB gain bandwidth is 11.15% from 27.1 to 30.3 GHz. The peak aperture efficiency can be calculated using the peak measured antenna gain at 28.2 GHz, and is found to be 31.1%. Some discrepancies are mainly due to practical fabrication tolerances and the waveguide-to-coaxial adapter effect (in the feed path) at the mmWave band.

The measured normalized radiation patterns at representative frequencies of 27, 28 and 30 GHz are presented in Fig. 10. Unanimously, the designed MFRA antenna exhibits stable, narrow beam broadside radiation patterns across the entire band of interest, and the measured patterns show excellent match with the simulated results in all cases. The measured HPBW is about 3.8° to 4°, SLLs are below −20 dB, and the X-pol levels are below −30 dB, as can be noticed directly from Fig. 10.

A performance comparison of the proposed MFRA with other related FRAs is presented in Table I. Notably, the proposed MFRA offers a wide impedance bandwidth, high gain, and a relatively low *H/D* ratio, along with the advantage of low-cost 3D-printed fabrication.

V. CONCLUSION

This work presented a fully 3D printed MFRA for high-gain FR-2 mmWave band applications. By exploiting control of volumetric unit-cell geometry, accurate phase compensation and efficient beam formation are achieved with a compact *H/D* ratio of 0.20. A prototype was manufactured through an in-house available low-cost 3D printing facility and measured, demonstrating a wide -10 dB impedance bandwidth covering 26 to 32 GHz. The proposed MFRA provides a peak gain of 31.1 dBi at 28.2 GHz, 11.5% 3-dB gain bandwidth centered at 28 GHz, and a peak aperture efficiency of 31.1%, with sidelobe and cross-polarization levels below -20 dB. An RMS prototype was also fabricated using an advanced commercial 3D-printing system. The performance of the proposed MFRA employing both the in-house and commercially manufactured RMS was experimentally measured and found to be in close agreement. The proposed additively manufactured MFRA provides a low-cost, robust, and high-performance antenna solution for high-gain millimeter-wave communication.